\documentclass[letterpaper]{article} 
\usepackage{aaai2026}  
\usepackage{times}  
\usepackage{helvet}  
\usepackage{courier}  
\usepackage[hyphens]{url}  
\usepackage{graphicx} 
\usepackage{natbib}  
\usepackage{caption} 
\usepackage{algorithm}
\usepackage{algorithmic}
\usepackage{multirow}
\usepackage{graphicx}
\usepackage{booktabs}
\usepackage{adjustbox}
\usepackage{amsmath}
\usepackage{array}
\usepackage{colortbl}
\usepackage{xcolor}
\usepackage{tcolorbox}
\usepackage{tabularx}
\usepackage{amssymb}

\usepackage{newfloat}
\usepackage{listings}
\DeclareCaptionStyle{ruled}{labelfont=normalfont,labelsep=colon,strut=off} 
\floatstyle{ruled}
\newfloat{listing}{tb}{lst}{}
\floatname{listing}{Listing}
\title{MicLog: Towards Accurate and Efficient LLM-based Log Parsing\\via Progressive Meta In-Context Learning}
\author{
    Jianbo Yu\textsuperscript{\rm 1},
    Yixuan Li\textsuperscript{\rm 2},
    Hai Xu\textsuperscript{\rm 3},
    Kang Xu\textsuperscript{\rm 4}, 
    Junjielong Xu\textsuperscript{\rm 5}, \\
    Zhijing Li\textsuperscript{\rm 6,7}, 
    Pinjia He\textsuperscript{\rm 5}, 
    Wanyuan Wang\textsuperscript{\rm 1}\thanks{Corresponding Author.}
}
\affiliations{

    \textsuperscript{\rm 1}School of Computer Science and Engineering, Southeast University, China\\
    \textsuperscript{\rm 2}College of Computing and Data Science, Nanyang Technological University, Singapore\\
    \textsuperscript{\rm 3}Focus Technology Co., Ltd., China\\
    \textsuperscript{\rm 4}School of Computer Science, Nanjing University of Posts and Telecommunications, China\\
    \textsuperscript{\rm 5}School of Data Science, The Chinese University of Hong Kong, Shenzhen (CUHK-Shenzhen), China\\
    \textsuperscript{\rm 6}School of Artificial Intelligence, Shenzhen University, China\\
    \textsuperscript{\rm 7}National Engineering Laboratory for Big Data System Computing Technology, Shenzhen University, China\\
    \{jianboyu,wywang\}@seu.edu.cn, yixuan.li@ntu.edu.sg, xuhai@focuschina.com, kxu@njupt.edu.cn, junjielongxu@link.cuhk.edu.cn, hepinjia@cuhk.edu.cn, lizhijing@szu.edu.cn

}

\usepackage{bibentry}

\begin{document}

\maketitle

\begin{abstract}
Log parsing converts semi-structured logs into structured templates, forming a critical foundation for downstream analysis. 
Traditional syntax and semantic-based parsers often struggle with semantic variations in evolving logs and data scarcity stemming from their limited domain coverage.
Recent large language model (LLM)-based parsers leverage in-context learning (ICL) to extract semantics from examples, demonstrating superior accuracy. However, LLM-based parsers face two main challenges: 1) underutilization of ICL capabilities, particularly in dynamic example selection and cross-domain generalization, leading to inconsistent performance; 2) time-consuming and costly LLM querying.
To address these challenges, we present MicLog, the first progressive meta in-context learning (ProgMeta-ICL) log parsing framework that combines meta-learning with ICL on small open-source LLMs (i.e., Qwen-2.5-3B). Specifically, MicLog: i) enhances LLMs' ICL capability through a zero-shot to k-shot ProgMeta-ICL paradigm, employing weighted DBSCAN candidate sampling and enhanced BM25 demonstration selection; ii) accelerates parsing via a multi-level pre-query cache that dynamically matches and refines recently parsed templates.
Evaluated on Loghub-2.0, MicLog achieves 10.3\% higher parsing accuracy than the state-of-the-art parser while reducing parsing time by 42.4\%. 
\end{abstract}


\section{Introduction}\label{sec:introduction}
\begin{figure*}
  \centering
  \includegraphics[width=\linewidth]{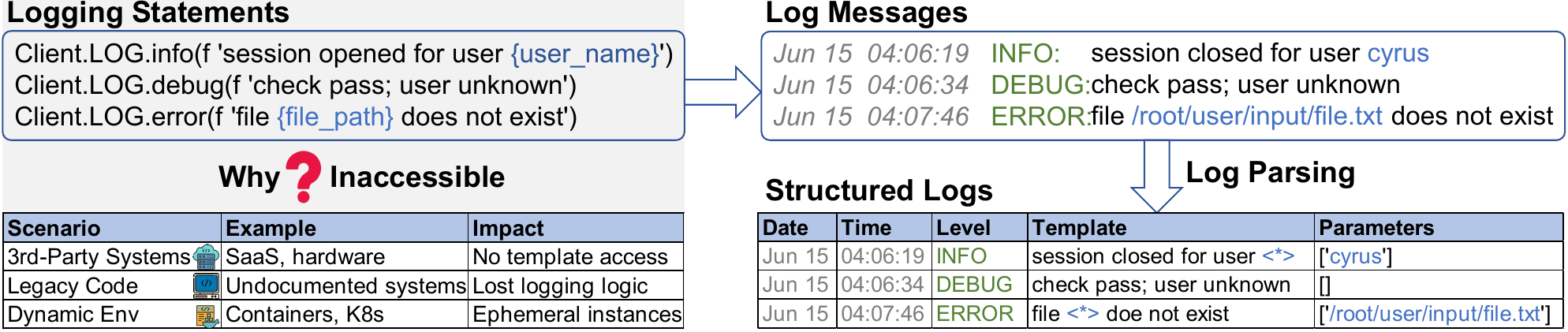}
  \caption{A simple process of Log Parsing.}
  \label{fig:log2fig}
\end{figure*}

In modern software systems, log data is crucial for maintenance and monitoring. These systems generate vast quantities of log messages, serving as indispensable resources for subsequent tasks including anomaly detection \cite{du2017deeplog, he2016experience, zhang2019robust}, root cause analysis \cite{amar2019mining, he2018identifying, lin2016log, huang2024demystifying}, and vulnerability prediction \cite{bilgin2020vulnerability, han2017learning}. However, the immense scale and complexity of logs make manual analysis impractical \cite{he2021survey}, necessitating automated log parsing techniques.
Log parsing, a fundamental step in log analysis, transforms messages into structured formats by extracting: 1) \textit{log templates} (consistent parts from logging statements), and 2) \textit{log parameters} (dynamic parts that vary per execution). As shown in Figure \ref{fig:log2fig}, the logging statement \texttt{client.LOG.info(f"session closed for user \{user\_name\}")} generates messages like \texttt{"session closed for user cyrus"}, where the template is \texttt{"session closed for user <*>"} and the parameter is \texttt{"cyrus"}.

Industrial systems, such as black-box microservices, IoT devices, and proprietary SaaS tools often produce logs with unknown formats, making parsing essential for reconstructing structures without access to the logging statements. Given the impracticality of accessing source code, various automated parsers have been developed:
(1) Syntax-based parsers \cite{dai2020logram, du2016spell, he2017drain, yu2023brain} use pattern recognition and clustering without prior format knowledge
(2) Semantic-based parsers \cite{huo2023semparser, le2023log, liu2022uniparser} leverage labeled data to train classification models
(3) Large language model (LLM) based parsers \cite{xu2024divlog, jiang2024lilac, xiao2024free, huang2024lunar, ma2024librelog, wu2024log} utilize LLMs to capture complex textual patterns.
These tools assist engineers in completing this critical initial step of log analysis.

While each log parsing approach has strengths, significant limitations persist. Syntax-based parsers struggle with highly variable formats, leading to inaccurate parsing. Semantic-based parsers require extensive labeled data and generalize poorly to unseen log formats. Although LLM-based parsers leverage in-context learning (ICL) for improved performance, recent studies \cite{le2023log, xu2024divlog} reveal they underutilize ICL capabilities for complex logs. Additionally, the time-consuming and costly nature of LLM querying, coupled with privacy concerns arising from leveraging commercial models like GPT to process sensitive log data, hinders their industrial deployment.

To address these limitations, we propose MicLog, an effective and efficient progressive meta in-context learning (ProgMeta-ICL) log parsing framework that leverages ProgMeta-ICL to enhance the ICL capabilities of LLMs in log parsing tasks. 
Specifically, MicLog comprises three main components: a weighted DBSCAN sampler, a progressive meta in-context training module, and a multi-level cache-enhanced LLM ICL log parser (MLCELI-Parser).
The weighted DBSCAN sampler employs a carefully designed algorithm to extract samples from open-source datasets based on a predefined sampling ratio after log preprocessing. These extracted samples serve two purposes:
(1) First, they are used for progressive meta in-context training to enhance ICL capabilities;
(2) Second, the MLCELI-Parser employs them to construct ICL prompts when no matching templates are found in the cache.
Before querying the LLM, MLCELI-Parser matches raw logs against cached templates and updates the cache with new templates derived from LLM query results to improve efficiency.

MicLog has been thoroughly evaluated on all 14 public datasets of Loghub-2.0 \cite{jiang2024large}. The results show that MicLog achieves the highest average accuracy on all performance metrics, achieving (1) 97.6\% Parsing Accuracy, (2) 95.3\% Precision Template Accuracy, and (3) 90.5\% Recall Template Accuracy when using same-source prompt examples. This outperforms the current state-of-the-art method AdaParser \cite{wu2024log} by 10.3\%, 12.6\%, and 6.1\%, respectively. 
Moreover, driven by its multi-level cache mechanism, MicLog achieves a 42.4\% reduction in total parsing time compared to AdaParser, while also surpassing the most efficient baseline Drain \cite{he2017drain}. The evaluation results demonstrate that MicLog is an effective and efficient log parsing framework in real-world deployment.

This paper presents the following key contributions:
\begin{itemize}
\item We propose MicLog, the first ProgMeta-ICL log parsing framework, which effectively addresses the limitations of existing LLM-based parsers.
\item We introduce a meta in-context training paradigm that facilitates efficient meta-learning, gradually transitioning from zero-shot to few-shot learning. This enables LLMs to improve their ICL capabilities.
\item We propose a multi-level cache-enhanced LLM ICL log parser (MLCELI-Parser) to matches input logs against cached templates and updates the cache with new templates derived from LLM ICL outputs when misses occur. 
\item We present the evaluation of MicLog on public datasets using three different performance evaluation metrics. The results show that MicLog achieves state-of-the-art performance, surpassing existing LLM-based parsers.
\end{itemize}

\section{Related Work}
\subsection{Log Parsing} \label{sec:log parsing}
Log parsing, the foundational phase of automated log analysis \cite{he2021survey, zhu2019tools}, extracts templates from raw messages by distinguishing variables from constants to produce structured logs. As Figure \ref{fig:log2fig} illustrates, parsers first extract headers (timestamps and verbosity levels) using regular expressions \cite{li2024revisiting} due to their predictable structure, with research primarily focusing on deriving templates and parameters from log message bodies. Industrial challenges arise from growing log volumes and evolving template complexity, where source code access would facilitate constant extraction but security and privacy constraints often prohibit this. Consequently, automated parsers have emerged—syntax-based \cite{shima2016length, dai2020logram, du2016spell, he2017drain, yu2023brain}, semantic-based \cite{huo2023semparser, le2023log, liu2022uniparser}, and LLM-based \cite{xu2024divlog, jiang2024lilac, xiao2024free, huang2024lunar, ma2024librelog, wu2024log, zhong2024logparser}—yet these exhibit limitations: syntax-based parsers rely on domain-specific features that may misalign with actual log content; semantic-based parsers struggle to generalize across highly diverse datasets; and LLM-based parsers, despite improved robustness, still underperform on complex datasets with inadequate robustness metrics. We therefore posit that fully leveraging LLMs' ICL potential is essential to enhance accuracy and robustness for complex log parsing tasks.
\begin{figure*}[h]
  \centering
  \includegraphics[width=\linewidth]{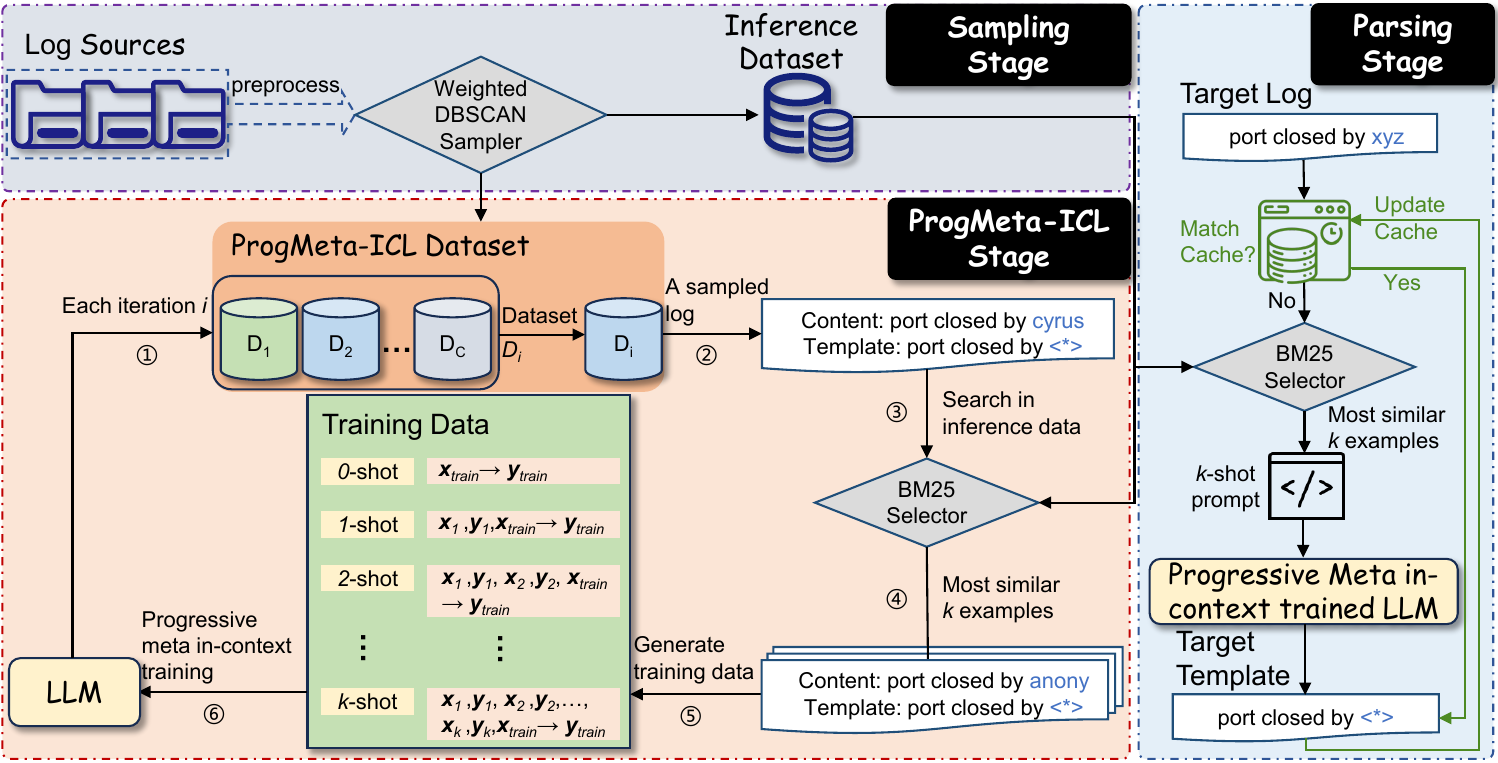}
  \caption{The workflow of MicLog framework. Sampling and ProgMeta-ICL are performed before online log parsing. The ProgMeta-ICL dataset and inference dataset need to be labeled before progressive meta in-context training and log parsing.}
  \label{fig:overview}
\end{figure*}
\subsection{Large Language Models and In-Context Learning}
LLMs \cite{zhao2023survey, yao2024survey}, trained on massive corpora via self-supervised learning, have revolutionized NLP with capabilities in text generation, language understanding, and complex reasoning, capturing rich semantic and syntactic patterns within their billion-parameter architectures. 
A key capability enabling their flexibility is ICL \cite{dong2022survey, min2022rethinking, akyurek2022learning, xu2024unilog, li2023revisiting}, which allows LLMs to adapt to tasks without fine-tuning. By providing few-shot examples directly within prompts, LLMs infer patterns and generate appropriate responses while leveraging pre-trained knowledge. This approach eliminates the need for extensive labeled training data and has proven effective across diverse NLP tasks including sentiment analysis, QA, and code generation.
For the critical automation task of log parsing, where traditional syntax or semantic-based approaches struggle with real-world complexity and diversity, LLMs with ICL offer a promising alternative by leveraging their natural language understanding to handle log semantics, adapting to new formats via example-based prompting, and operating effectively with minimal labeled data. When provided with log-template examples in prompts, LLMs can accurately parse unseen logs, an adaptability particularly valuable for heterogeneous systems with significantly varying log structures, as it eliminates the need for retraining.

\subsection{Meta In-Context Learning}
While ICL enables LLMs to adapt to tasks using few-shot examples in prompts, its effectiveness diminishes when task inputs are low-quality or inherently complex. Recent studies \cite{min2021metaicl, chen2021meta, li2024meta, coda2023meta} address this through meta in-context learning (Meta-ICL), which trains models explicitly to enhance ICL capabilities via multitask learning using fixed-shot examples. However, existing methods rely on fixed example counts per task during meta-training, which is suboptimal as tasks vary in their example requirements.
To address the limitations of Meta-ICL, we extend this paradigm with ProgMeta-ICL: a flexible, dynamic approach featuring a two-stage training process. First, the model is trained across diverse tasks under few-shot conditions to capture foundational patterns. Subsequently, examples per log parsing task are progressively increased to refine task-specific understanding. This staged exposure enables adaptive learning, enhances complex task performance and improves robustness across diverse scenarios.

\section{Method}
In this section, we introduce MicLog, a ProgMeta-ICL LLM-based log parsing framework. As illustrated in Figure \ref{fig:overview}, MicLog is composed of three components, and its overall procedure is summarized in Algorithm 1.
\begin{table*}[ht]
\centering
\small 
\begin{tabular}{m{2.5cm}|p{13.5cm}}  
\toprule
 & \textbf{Task:} $C$ progressive meta in-context training tasks \\ 
\cmidrule(l){2-2} 
                                  & \textbf{Data given:} Training examples $T_i = \{(x_j^i, y_j^i)\}_{j=1}^{N_i}, \forall i \in [1, C] (N_i \gg K)$, max shot number $K$ \\ 
\cmidrule(l){2-2} 
\centering \textbf{Progressive meta in-context training} & \textbf{Objective:} \begin{tabular}[t]{@{}l@{}} For each iteration,\\ 1. Sample task $i \in [1, C]$ \\ 2. Iterate over shot number $k$ from $0$ to $K$ \\ 3. Sample $k$ examples from $T_i$: $(x_1, y_1), \cdots, (x_{k}, y_{k})$ \\ 4. Maximize $P(y_{train}|x_1, y_1, \cdots, x_k, y_k, x_{train})$ \end{tabular} \\ 
\midrule
     & \textbf{Task:} An unseen \textit{target} task \\ 
\cmidrule(l){2-2} 
\centering \textbf{Inference}     & \textbf{Data given:} Training examples $(x_1, y_1), \cdots, (x_k, y_k)$, Test input $x_{target}$ \\ 
\cmidrule(l){2-2} 
                                  & \textbf{Objective:} $\arg\max_{c \in \mathcal{C}} P(c|x_1, y_1, \cdots, x_k, y_k, x_{target})$ \\ 
\bottomrule
\end{tabular}
\caption{Progressive meta in-context training and inference process}
\label{tab:meta_training_inference}
\end{table*}
\begin{algorithm}[tb]
\small
\caption{MicLog: Progressive Meta ICL Log Parsing}
\label{alg:framework}
\begin{algorithmic}[1]
\REQUIRE 
  Raw log dataset $\mathcal{D}$, Initial LLM parameters $\theta$, Max shot number $k$, Sampling ratio $\alpha$, LRU cache capacity $\mathit{Cache}_{\text{LRU}}$, Pattern cache $\mathit{Cache}_{\text{Pattern}}$
\ENSURE 
  Parsed templates $\mathcal{T}$, Meta-trained LLM parameters $\theta^*$

    \STATE \textbf{Stage 1: Sampling} 
    \STATE $\mathcal{D}_{\text{dedup}} \gets \text{deduplicate}(\mathcal{D})$
    \STATE $\mathcal{S}_{\text{meta}}, \mathcal{S}_{\text{inf}} \gets \text{WeightedDBSCAN}(\mathcal{D}_{\text{dedup}}, \alpha)$ 
    
    \STATE \textbf{Stage 2: ProgMeta-ICL} 
      \FOR{$shot = 0$ \TO $k$} 
        \STATE Sample task batch $\mathcal{B} \sim \mathcal{S}_{\text{meta}}$
        \STATE Concatenate $shot$ examples: $\mathit{prompt} \gets \{(x_j,y_j)\}_{j=1}^{shot}$
        \STATE Update $\theta \gets \theta - \nabla \mathcal{L}_{\text{ProgMeta-ICL}}$ 
      \ENDFOR
    \STATE $\theta^* \gets \theta$
    
    \STATE \textbf{Stage 3: Parsing with Multi-Level Cache}
    \FOR{each raw log $l_i \in \mathcal{D}$}
      \IF{$l_i \in \mathit{Cache}_{\text{LRU}}$} 
        \STATE $t_i \gets \mathit{Cache}_{\text{LRU}}[l_i]$
      \ELSIF{$\exists t_p \in \mathit{Cache}_{\text{Pattern}}$ s.t. $\text{validate}(l_i, t_p)$} 
        \STATE $t_i \gets t_p$, Update $\mathit{Cache}_{\text{LRU}}$ with $(l_i, t_i)$
      \ELSE 
        \STATE Retrieve top-$k$ logs $\mathcal{R} \gets \text{BM25}(l_i, \mathcal{S}_{\text{inf}})$ 
        \STATE Construct ICL prompt $P \gets \{\mathcal{R}, l_i\}$, $t_i \gets \text{LLM}_{\theta^*}(P)$ 
        \STATE Update $\mathit{Cache}_{\text{LRU}}$, $\mathit{Cache}_{\text{Pattern}}$ with $(l_i, t_i)$
      \ENDIF
      \STATE $\mathcal{T} \gets \mathcal{T} \cup \{t_i\}$
    \ENDFOR
\end{algorithmic}
\end{algorithm}

\subsection{Weighted DBSCAN Sampler}
Both the ProgMeta-ICL and parsing stages begin by sampling a small subset of candidate log-template pairs. This subset must be diverse to avoid LLM overfitting and sufficiently representative to cover varied logs and their key characteristics.

After deduplication in preprocess, we propose a weighted DBSCAN sampling method to extract small, diverse, and representative log subsets. DBSCAN \cite{ester1996density, schubert2017dbscan} clusters densely packed points using parameters $\epsilon$ (neighborhood radius) and $MinPts$ (minimum cluster density), labeling sparse points as noise. We leverage DBSCAN's core points for clustering and incorporate weighting to efficiently sample representative logs. For each log $l \in \mathcal{D}$, complexity is computed via Equation \ref{eq:log_complexity}.

\begin{equation}
\label{eq:log_complexity}
\textit{complexity(l)} = \textit{${token_{l}}^{token_{l}}$} + \textit{$length_{l}$}
\end{equation}
Here, $token_{l}$ equals to the number of token of log $l$ and $length_l$ means log's length. To prevent numerical instability, we introduce a constant smoothing factor $factor_s$ for computing weights that incorporate each log's complexity.
\begin{equation}
\label{eq:weights}
\textit{w} = \frac{\textit{complexity}(l) + factor_s}{\sum_{l_i \in \mathcal{D}} [\textit{complexity}(l_i) + factor_s]}
\end{equation}

DBSCAN also introduces neighborhood radius and minimum points to assist in completing the clustering process. Besides, the complexity is also utilized here to calculate the Euclidean distance of the complexity of a log. After clustering concludes, samples are extracted from each cluster in accordance with the weight computed using Equation \ref{eq:weights}, based on the specified sample ratio. We reuse this algorithm and obtained two datasets, one is ProgMeta-ICL dataset to complete meta in-context training, and the other is the inference dataset for generating ICL prompt.
\subsection{Progressive Meta In-Context Training}
Table \ref{tab:meta_training_inference} provides an overview of the progressive meta in-context training process in MicLog. The key idea is to use a multi-task learning scheme over a large collection of meta-training tasks, in order for the LLMs to learn how to condition on a small set of ICL examples, understand the core content of the current task from it, and provide the corresponding log template output. The meta in-context training examples are concatenated as a single input to the LLM, which sequentially presents the 0-shot to $k$-shot (e.g., $k$ = 5) learning procedure. At test time, the meta in-context trained LLM is evaluated on unseen target tasks that come with $k$ training examples, and inference directly follows the same data format as in meta-training.

By analyzing the work of \textit{Zhang et al.} \cite{zhang2023and}, in the ProgMeta-ICL stage, the goal is to minimize the meta-loss across a range of tasks $\mathcal{T}$, where each task $\mathcal{T}_i$ requires its own ICL setting. Meta-learning aims to train a model that can rapidly adapt to new tasks by leveraging prior learning. For each task $\mathcal{T}_i$ drawn from a meta-distribution $\mathcal{P}(\mathcal{T})$, let $S_T^{(\mathcal{T}_i)}$ denotes the demonstration set for task $\mathcal{T}_i$, and the meta-objective is to minimize the expected ICL loss over tasks:
\begin{equation}
\label{eq:ProgMeta-ICL}
\mathcal{L}_{ProgMeta-ICL} = \mathbb{E}_{\mathcal{T} \sim \mathcal{P}(\mathcal{T})} \left[ \mathcal{L}_{ICL}(\mathcal{T}, S_T^{(\mathcal{T})}) \right]  
\end{equation}

The meta-learner aims to minimize this loss across tasks by learning a shared representation $\theta$. For a meta trained LLM, when facing any new task $\mathcal{T}_j$ by leveraging ICL paradigm, the model can quickly adapt to minimize:
\begin{equation}
\label{eq:ICL}
\mathcal{L}_{ICL}(\mathcal{T}_j) = \frac{1}{T} \sum_{t=1}^{T} - \log P(r_t|x_t, S_{t-1}, \theta)
\end{equation}
where \( r_t \) represents the verbalizer associated with input \( x_t \) for task \( \mathcal{T}_i \), \( x_t \) is the input at step \( t \), \( S_{t-1} \) is the set of input-label pairs up to step \( t-1 \). The parameter \( \theta \) is optimized by the meta-learner to be task-agnostic and it can quickly adapt for different tasks. 

\subsection{Multi-level Cache Enhanced LLM ICL Log Parser}
MicLog employs a multi-level caching mechanism to minimize redundant LLM invocations and accelerate parsing by exploiting structural patterns in log streams. The cache architecture consists of two synergistic components:

\begin{itemize}
    \item \textbf{LRU Cache:} Maintains recently parsed $\langle$raw log, template$\rangle$ pairs in an ordered dictionary with fixed capacity $Cache_{\text{LRU}}$. Entries satisfy $\|l_i\|_{\text{tokens}} \leq \tau$, with least-recently-used eviction when exceeding capacity:
    \begin{equation}
        \text{Cache}_{\text{LRU}} = \{(l_i, t_i) \mid i \in [1, C_{\text{LRU}}]\}
    \end{equation}
    
    \item \textbf{Pattern Cache:} Stores template patterns $t_p$ for structural matching, where $\eta(\cdot)$ performs pattern normalization:
    \begin{equation}
        \text{Cache}_{\text{Pattern}} = \{t_p \mid t_p = \eta(t_{\text{parsed}})\}
    \end{equation}
\end{itemize}

Cache lookup follows a two-stage process. First, the raw log $l_{\text{raw}}$ is checked against the LRU cache for an exact match. If found, the corresponding template is returned immediately. On LRU miss, $l_{\text{raw}}$ is normalized and compared against all patterns in the pattern cache. The validation function $\text{validate}(l_{\text{norm}}, t_p)$ decomposes $t_p$ into constant segments $\{s_j\}$ separated by \texttt{"<*>"}, verifying their ordered occurrence in $l_{\text{norm}}$ with strictly increasing positions $\text{pos}_j$:
\begin{equation}
    \forall s_j \in \text{segments}(t_p),\ \exists\ \text{pos}_j \mid l_{\text{norm}}[\text{pos}_j : \text{pos}_j + \|s_j\|] = s_j
\end{equation}

This design achieves $\mathcal{O}(1)$ exact-match lookups and $\mathcal{O}(k)$ pattern-match lookups ($k = \|\text{Cache}_{\text{Pattern}}\|$). Upon successful pattern match, the template $t_p$ is cached in the LRU and returned. On full cache miss, newly generated templates, which returned by the LLM ICL log parser, update both caches.

Our LLM-based log parser begins with constructing effective prompts. We select $k$ semantically relevant logs from the inference dataset. Studies demonstrate that example selection critically impacts LLM performance \cite{rubin2021learning}, with recent work \cite{xu2024divlog} showing that ordering $k$-shot examples by ascending similarity maximizes log parsing accuracy. 

We implement an enhanced BM25 algorithm \cite{robertson2009probabilistic} for efficient similarity search. For a log corpus $\mathcal{D}$ containing $N$ entries, we compute the Inverse Document Frequency (IDF) \cite{robertson2004understanding} for each $w$ as:
\begin{equation}
\label{eq:IDF}
\text{IDF}(w) = \log\left(\frac{N - f(w) + 0.5}{f(w) + 0.5} + 1\right)
\end{equation}
where $f(w)$ denotes the document frequency of $w$. The BM25 score between query log $q$ and candidate log $d$ is:
\begin{equation}
\label{eq:BM25}
\text{BM25} = \sum_{w \in q} \text{IDF}(w) \cdot \frac{\text{TF}(w, d) \cdot (k_1 + 1)}{\text{TF}(w, d) + k_1 \cdot \left(1 - b + b \cdot \frac{|d|}{\text{avg\_l}}\right)}
\end{equation}

where $\text{TF}(w,d)$ is $w$'s term frequency in $d$, $|d|$ is log length, $avg\_l$ is average log length in $\mathcal{D}$, and $k_1$, $b$ are tunable parameters controlling term frequency saturation and length normalization. 

The top $k$ candidates identified by BM25 are sorted in descending similarity order to generate the prompt. The parsed log template for the current raw log is retrieved via the meta in-context trained LLM query.

\section{Evaluation}
In this section, we outline the experimental setup, followed by the evaluation results and analysis conducted on public datasets to address the following research questions:

\noindent\textbf{RQ1:} How effective and stable is MicLog?

\noindent\textbf{RQ2:} How does each component contribute to MicLog?

\noindent\textbf{RQ4:} How efficient is MicLog?

\noindent\textbf{RQ4:} How do different training strategies affect MicLog?

\subsection{Experiment Setup}
\subsubsection{Datasets.}
Our experiments utilize Loghub-2.0 \cite{jiang2024large}, a comprehensive log parsing datasets from LogPAI \cite{zhu2019tools}. 
Loghub-2.0 contains 14 datasets of system logs from diverse sources such as distributed systems, supercomputer systems, and server-side applications, totaling over 50 million log messages and 3,488 log templates.
\subsubsection{Baselines.}\label{sec:baselines} 
We select Drain \cite{he2017drain}, Brain \cite{yu2023brain}, LogPPT \cite{le2023log}, LUNAR \cite{huang2024lunar}, LibreLog \cite{ma2024librelog}, LILAC \cite{jiang2024lilac} and AdaParser \cite{wu2024log} as our baselines for comparison. The first two parsers demonstrate superior performance among all syntax-based parsers. The semantic-based parser, LogPPT leverages template-free prompt-tuning \cite{ma2021template} to fine-tune a pre-trained language model, RoBERTa \cite{liu2019roberta}. LLM-based paser LUNAR leverages log contrastive units to facilitate effective comparisons by the LLM. LibreLog uses open-source LLMs to parse logs by syntactic similarity in the static text. LILAC utilizes the ICL capability to adapt LLMs to parse various log data with adaptive parsing cache. AdaParser improves parsing accuracy based on SG-ICL and selfcorrection.
Due to the unavailability of the original GPT-3 versions \cite{brown2020language} required by LUNAR, LibreLog, LILAC and AdaParser, we replicate their experiments using \textit{gpt-3.5-turbo-0125} for comparative analysis.

\begin{table*}[htbp]
\centering
\resizebox{\textwidth}{!}{%
\small
\setlength{\tabcolsep}{2pt} 
\begin{tabular}{l | c c c | c c c | c c c | c c c | c c c | c c c | c c c | c c c}
\toprule
\multirow{2}{*}{\textbf{Dataset}} & \multicolumn{3}{c|}{\textbf{Drain}} & \multicolumn{3}{c|}{\textbf{Brain}} & \multicolumn{3}{c|}{\textbf{LogPPT}} & \multicolumn{3}{c|}{\textbf{LUNAR}} & \multicolumn{3}{c|}{\textbf{LibreLog}} & \multicolumn{3}{c|}{\textbf{LILAC}} & \multicolumn{3}{c|}{\textbf{AdaParser}} & \multicolumn{3}{c}{\textbf{MicLog}}  \\ 
\cmidrule(lr){2-4} \cmidrule(lr){5-7} \cmidrule(lr){8-10} \cmidrule(lr){11-13} \cmidrule(lr){14-16} \cmidrule(lr){17-19} \cmidrule(lr){20-22} \cmidrule(lr){23-25}
 & PA & PTA & RTA & PA & PTA & RTA & PA & PTA & RTA & PA & PTA & RTA & PA & PTA & RTA & PA & PTA & RTA & PA & PTA & RTA & PA & PTA & RTA \\
\midrule
Apache & 72.3 & 48.3 & 50.0 & 28.7 & 45.2 & 50.0 & 94.8 & 39.8 & 34.6 & 50.8 & 69.0 & 69.0 & 99.6 & 85.4 & 92.9 & 99.5 & 82.8 & 82.8 & 99.9 & 93.1 & 93.1 & \textbf{100.0} & \textbf{100.0} & \textbf{100.0} \\
BGL & 40.7 & 16.4 & 21.3 & 40.2 & 15.9 & 22.5 & 93.8 & 30.5 & 22.4 & 60.6 & 79.0 & 82.2 & 92.9 & 86.4 & 84.1 & 95.3 & 73.9 & 80.6 & 98.0 & 80.8 & 79.1 & \textbf{99.0} & \textbf{92.5} & \textbf{85.3} \\
Hadoop & 51.2 & 32.0 & 46.0 & 14.1 & 15.2 & 29.5 & 66.6 & 49.6 & 38.4 & 85.9 & 73.4 & 75.8 & 87.1 & 87.9 & 87.7 & 83.9 & 77.2 & 75.8 & \textbf{95.5} & 83.4 & 85.2 & 94.0 & \textbf{97.9} & \textbf{96.6} \\
HDFS & 62.1 & 58.7 & 58.7 & 92.9 & 63.4 & 56.5 & 94.3 & 28.0 & 35.2 & 87.4 & 97.8 & 97.8 & \textbf{100.0} & 98.4 & 98.4 & \textbf{100.0} & 95.7 & 97.8 & \textbf{100.0} & \textbf{100.0} & \textbf{100.0} & \textbf{100.0} & \textbf{100.0} & \textbf{100.0} \\
HealthApp & 18.3 & 0.2 & 36.6 & 17.1 & 29.4 & 35.8 & 99.7 & 82.2 & 82.2 & 98.2 & 86.9 & \textbf{89.1} & 97.4 & 87.9 & 90.4 & 73.6 & 85.0 & 87.2 & 86.4 & 88.5 & \textbf{89.1} & \textbf{100.0} & \textbf{93.9} & 88.5 \\
HPC & 72.1 & 9.0 & 46.8 & 66.3 & 14.4 & 48.4 & 99.7 & 70.8 & 81.4 & 99.0 & 69.7 & 93.2 & 97.3 & 88.4 & 81.9 & 70.3 & 77.3 & 78.4 & 80.3 & \textbf{89.2} & \textbf{89.2} & \textbf{99.6} & 80.0 & 54.1 \\
Linux & 9.9 & 17.7 & 25.3 & 8.7 & 21.1 & 29.2 & 16.8 & 26.8 & 52.3 & 84.3 & 72.6 & 76.0 & 90.2 & 89.6 & 90.4 & 84.2 & 73.9 & 73.7 & 73.2 & 75.0 & 76.3 & \textbf{99.1} & \textbf{96.6} & \textbf{92.9} \\
Mac & 28.1 & 3.8 & 25.2 & 32.4 & 24.9 & 33.4 & 39.0 & 25.6 & 32.6 & 59.2 & 55.4 & 56.5 & 65.4 & 50.2 & 60.5 & 68.5 & 54.0 & 64.1 & 57.9 & 59.4 & 58.8 & \textbf{95.4} & \textbf{96.4} & \textbf{94.4} \\
OpenSSH & 58.5 & 37.5 & 42.9 & 48.1 & 26.5 & 37.1 & 65.4 & 8.1 & 14.3 & 69.1 & 80.5 & 86.8 & 49.6 & 34.8 & 30.6 & \textbf{96.8} & 85.7 & 78.9 & 94.2 & 92.3 & \textbf{94.7} & 94.2 & \textbf{94.6} & 92.1 \\
OpenStack & 2.9 & 0.1 & 14.6 & 14.1 & 29.2 & 29.2 & 40.6 & 70.4 & 78.3 & 94.2 & 82.4 & 87.5 & 83.1 & 72.9 & 68.2 & \textbf{100.0} & 93.8 & 93.8 & \textbf{100.0} & 97.9 & 97.9 & \textbf{100.0} & \textbf{100.0} & \textbf{100.0} \\
Proxifier & 68.8 & 8.8 & 45.5 & 70.3 & 87.5 & 63.6 & \textbf{100.0} & 95.7 & 95.7 & 51.1 & 72.7 & 72.7 & 89.7 & 90.5 & 86.1 & \textbf{100.0} & 90.9 & 90.9 & 98.4 & 91.7 & \textbf{100.0} & \textbf{100.0} & \textbf{100.0} & \textbf{100.0} \\
Spark & 39.4 & 38.8 & 42.6 & 39.3 & 5.5 & 38.7 & 95.2 & 36.0 & 27.8 & 97.0 & 70.1 & 74.6 & 88.9 & 70.5 & 62.7 & 97.3 & 80.4 & 78.4 & 98.2 & 80.3 & 79.2 & \textbf{100.0} & \textbf{93.8} & \textbf{90.3} \\
Thunderbird & 21.6 & 4.0 & 26.6 & 26.1 & 23.7 & 30.3 & 40.1 & 13.5 & 9.2 & 63.5 & 60.7 & 65.6 & 69.4 & 52.0 & 48.6 & 62.6 & 46.4 & 64.1 & 64.4 & 62.3 & 63.1 & \textbf{85.6} & \textbf{92.9} & \textbf{81.6} \\
Zookeeper & 84.3 & 64.9 & 57.5 & 82.2 & 57.4 & 62.1 & 84.5 & 80.9 & 80.9 & 71.2 & 88.3 & 76.4 & 85.0 & 80.1 & 78.9 & 68.5 & 89.7 & 87.6 & 92.0 & 90.8 & 88.8 & \textbf{99.3} & \textbf{95.3} & \textbf{91.0} \\
\midrule
\textbf{Average} & 45.0 & 24.3 & 38.5 & 41.5 & 32.8 & 40.5 & 73.6 & 47.0 & 49.0 & 76.5 & 75.6 & 78.8 & 85.4 & 76.8 & 75.8 & 85.8 & 79.1 & 81.0 & 88.5 & 84.6 & 85.3 & \textbf{97.6} & \textbf{95.3} & \textbf{90.5} \\
\bottomrule
\end{tabular}
}
\caption{Performance of Various Log Parsing Methods on Loghub-2.0 Datasets (\%)}
\label{tab:accuracy}
\end{table*}

\begin{figure*}[h]
  \centering
  \includegraphics[width=\linewidth]{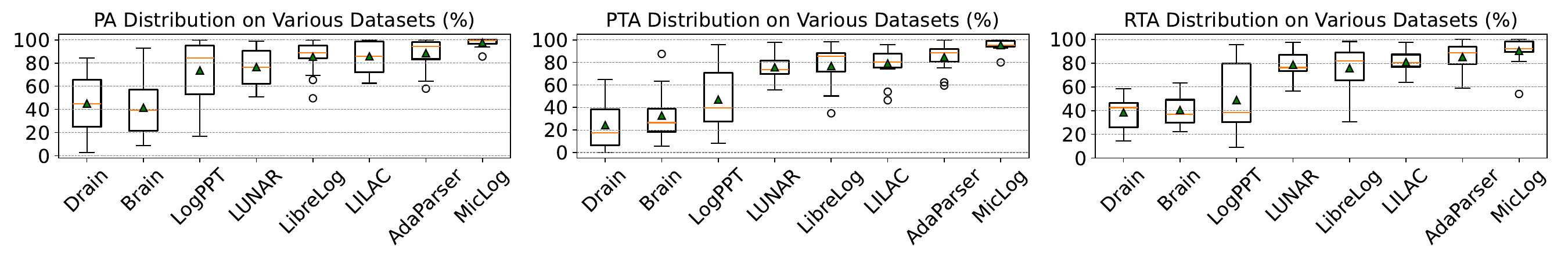}
  \caption{Robustness comparison between baselines and MicLog on public datasets (\%)}
  \label{fig:accuracy}
\end{figure*}

\subsubsection{Metrics.}
In line with recent studies \cite{khan2022guidelines, xu2024divlog}, our evaluation employs three metrics: Parsing Accuracy (PA), Precision Template Accuracy (PTA), and Recall Template Accuracy (RTA), with the latter two metrics collectively referred to as Template Accuracy (TA). The definitions of these metrics are as follows:
\begin{itemize}
\item Parsing Accuracy (PA) is used to evaluate the ability to correctly extract log templates, defined as the ratio of correctly parsed log messages to the total number of logs. 
\item Template Accuracy (TA) is a template-level metric calculated based on the proportion of correctly identified templates. Using the number of correctly identified templates ($N_c$), identified templates ($N_i$) and ground-truth templates ($N_g$), we calculate the Precision ($PTA=\frac{N_c}{N_i}$) and Recall ($RTA=\frac{N_c}{N_g}$). 
\end{itemize}

\subsubsection{Environment and Implementation.}
We conduct experiments using the local open-source LLM Qwen2.5-3B \cite{qwen2} for meta-learning and ICL inference on an RTX 4090 GPU to obtain raw prompt responses. We also employed Python 3.10 to implement the weighted DBSCAN sampler, BM25 selector, multi-level cache, and evaluation scripts on Ubuntu 22.04 LTS server. For meta in-context training, MicLog samples only 0.009\% (326 logs avg.) logs from Loghub-2.0 for maximum training efficiency. During ICL inference, 5 labeled examples per query are selected from the candidate set for prompt construction. We repeated each experimental configuration five times and calculated their mean as the final result. All configurations remain fixed throughout evaluation. 

\subsection{RQ1: Effectiveness}\label{sec:rq1}
In this section, we perform a comprehensive evaluation of the accuracy and robustness of MicLog, along with the baselines, across all datasets from Loghub-2.0. We assess the accuracy of the parsers using PA, PTA, and RTA, and evaluate their robustness by analyzing the statistical distribution. 

\subsubsection{Accuracy.}
Table \ref{tab:accuracy} presents comprehensive evaluation results on Loghub-2.0, with best metrics per dataset in \textbf{bold}. MicLog averages 97.6\% PA, 95.3\% PTA, and 90.5\% RTA, outperforming the SOTA pasers AdaParser by 10.3\%, 12.6\%, and 6.1\%, respectively. 
To assess if MicLog significantly outperforms AdaParser, we performed one-sided Wilcoxon signed-rank tests \cite{rey2011wilcoxon} for the three metrics across 14 datasets. 
\begin{table}[h]
\centering
\small 
\setlength{\tabcolsep}{3pt} 
\begin{tabular*}{\columnwidth}{@{\extracolsep{\fill}}l *{3}{>{\centering\arraybackslash}p{0.2\columnwidth}}}
\toprule
\textbf{Test Condition} & \textbf{PA} & \textbf{PTA} & \textbf{RTA} \\
\midrule
MicLog \textit{epoch=5} & 97.6 & 95.3 & 90.5 \\
MicLog \textit{epoch=1} & 96.3 (↓1.3) & 93.9 (↓1.4) & 89.7 (↓0.8) \\
w/o ProgMeta-ICL & 87.3 (↓10.3) & 60.2 (↓35.1) & 62.0 (↓28.5) \\
w/ k-means sampling & 97.1 (↓0.5) & 92.0 (↓3.3) & 88.5 (↓2.0) \\
w/ random sampling & 93.7 (↓3.9) & 76.8 (↓18.5) & 72.9 (↓17.4) \\
w/ LILAC cache & 97.6 (-) & 94.4 (↓0.9) & 89.0 (↓1.5) \\
w/ random selection & 94.7 (↓2.9) & 79.1 (↓16.2) & 73.4 (↓17.1) \\
w/ Proxifier only & 95.8 (↓1.8) & 90.2 (↓5.1) & 87.1 (↓3.4) \\
\bottomrule
\end{tabular*}
\caption{Average accuracy comparison on Loghub among MicLog with different strategies (\%)}
\label{tab:ablationtable}
\end{table}
The $p$-values (0.0038, 0.0036, 0.0356) are all below 0.05 and test statistics are (63.0, 84.0, 62.0), providing strong evidence for MicLog's superior performance.

\subsubsection{Robustness.} \label{sec:robustness}
Following recent work \cite{xu2024divlog}, we compare the robustness of MicLog and baselines using box plots of three metrics across datasets. The orange line marks the median and the green triangle marks the mean. As shown in Figure \ref{fig:accuracy}, MicLog attains the highest accuracy and the smallest variance, reflected by its narrowest distribution. This demonstrates MicLog’s strong robustness.

\subsection{RQ2: Ablation Study}\label{sec:rq2}

In this section, we conduct ablation study to discuss the contribution of each component in MicLog. 
\begin{figure}[h]
  \centering
  \includegraphics[width=\linewidth]{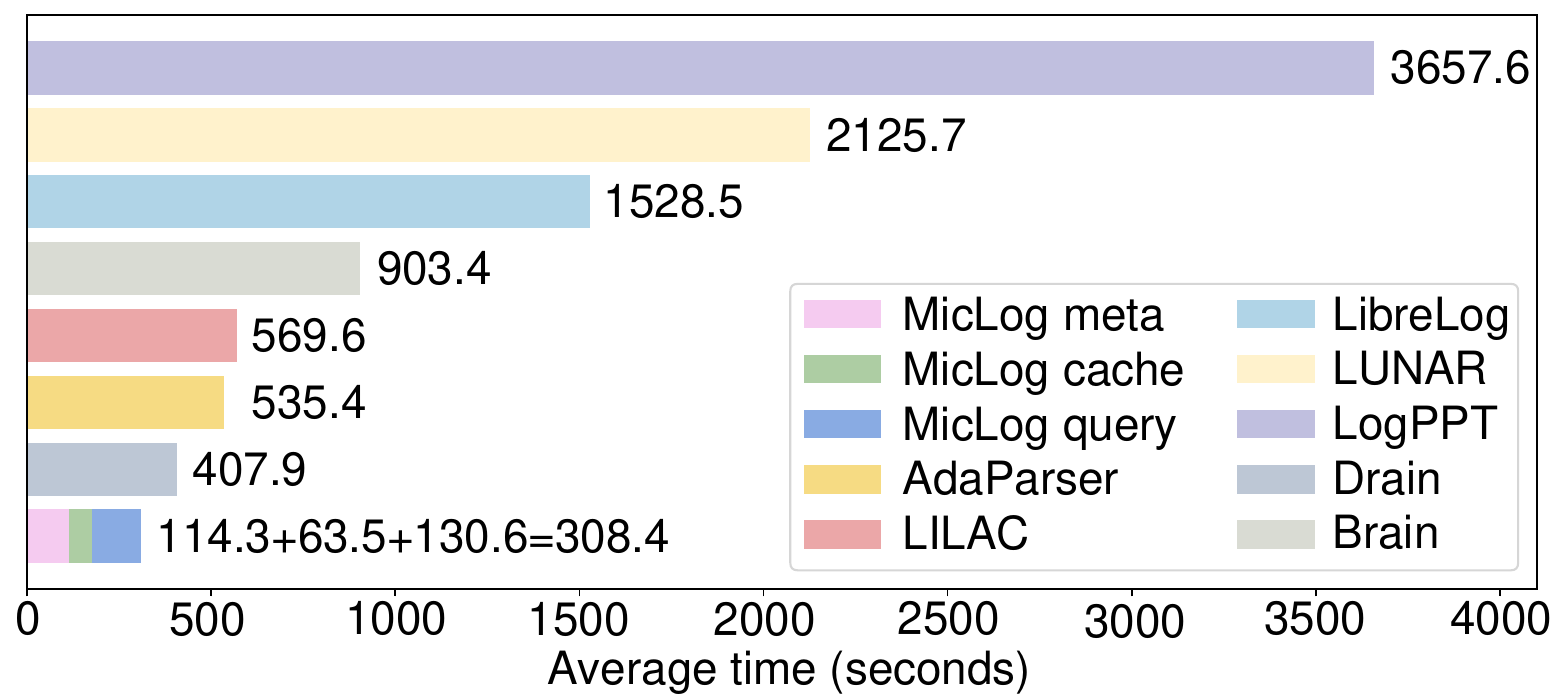}
  \caption{Efficiency of MicLog and baselines on Loghub-2.0}
  \label{fig:efficiency}
\end{figure}
We have developed seven variants of MicLog, the first one reset the hyperparameter \textit{epoch}, five of them were replaced based on the components of MicLog, and the last one was studied based on 1/14 of the training data, namely: 1) MicLog \textit{epoch=1}: reset epoch from 5 to 1; 2) MicLog w/o ProgMeta-ICL: remove the ProgMeta-ICL design; 3) MicLog w/ k-means sampling: replace the weighted DBSCAN sampler by a k-means sampler; 4) MicLog w/ random sampling: replace the weighted DBSCAN sampler by a random sampler; 5) MicLog w/ LILAC cache: replace the multi-level cache by cache in LILAC; 6) MicLog w/ random selection: replace the BM25 selector by a random selector; 7) MicLog w/ Proxifier only: employ meta in-context training only in dataset Proxifier and test on all datasets.

The detailed evaluation results are shown in Table. \ref{tab:ablationtable}, in which the following points can be made: 
(1) Progressive meta in-context train only 1 \textit{epoch} leads to good enough result to surpass the SOTA method.
(2) Removal of the ProgMeta-ICL module causes the most significant performance degradation, confirming its critical role in enhancing the LLM's in-context learning capability.
(3) Substituting the weighted DBSCAN sampler with k-means yields marginal performance drops, while random sampling results in substantially larger reductions. 
(4) Caching hardly affects MicLog's accuracy, since this component is designed to optimize efficiency (discussed in RQ3).
(5) Replacing the BM25 selector with random selection severely impacts average performance. 
(6) Generalization of ProgMeta-ICL paradigm: Proxifier represents a distinct distribution from other datasets. Our framework significantly enhances LLMs' ICL capabilities for log parsing using only Proxifier, achieving performance comparable to MicLog. This confirms our method genuinely improves ICL capabilities rather than dataset familiarity, and demonstrates effective generalization to diverse datasets with minimal training data.
\subsection{RQ3: Efficiency}\label{sec:rq3}
We evaluate MicLog and baselines on Logpub-2.0, recording average parsing times across datasets. 
The efficiency results are demonstrated in the Figure \ref{fig:efficiency}.
The time cost analysis for MicLog incorporates both meta in-context  training and parsing phase operations (cache retrieval and LLM queries). Results demonstrate that MicLog achieves lower average times across all Loghub-2.0 datasets than any baselines, 
\begin{table}[htbp]  
\centering  
\small
\setlength{\tabcolsep}{3pt} 
\begin{tabularx}{\columnwidth}{l *{3}{>{\centering\arraybackslash}X}}   
\toprule  
\textbf{Test Condition} & \textbf{PA} & \textbf{PTA} & \textbf{RTA} \\  
\midrule  
MicLog & 97.6 & 95.3 & 90.5 \\    
w/ 0-shot only & 94.4 (↓3.2) & 85.4 (↓9.9) & 79.4 (↓11.1) \\
w/ 1-shot only & 94.6 (↓3.0) & 86.8 (↓8.5) & 79.7 (↓10.8) \\
w/ 3-shot only & 94.6 (↓3.0) & 85.7 (↓9.6) & 80.1(↓10.4) \\
w/ 5-shot only & 95.8 (↓1.8) & 87.5 (↓7.8) & 81.4 (↓9.1) \\
w/o ProgMeta-ICL & 87.3 (↓10.3) & 60.2 (↓35.1) & 62.0 (↓28.5) \\ 
\bottomrule  
\end{tabularx} 
\caption{Average accuracy comparison among MicLog with different shot meta in-context learning strategies (\%)} 
\label{tab:shotdiff}
\end{table} 
reducing total parsing time by 42.4\% compared to AdaParser while outperforming even the most efficient baseline, Drain. Notably, MicLog's average cache time (63.5s) is substantially lower than LILAC's 376.5s (not list in the figure) despite both employing caching mechanisms. This efficiency stems from MicLog's multi-level architecture, which capitalizes on the temporal locality of log data - where identical or similar log entries frequently recur in short intervals. Our design significantly increases cache hit rates in such scenarios, accelerating parsing throughput.
Moreover, by leveraging the open-source Qwen-3B LLM, our approach effectively addresses privacy concerns while achieving substantially lower training and inference costs than methods relying on proprietary models like the GPT-3.5 series or resource-intensive alternatives such as Llama3-8B.

\subsection{RQ4: Impact of Different Training Stratgies}\label{sec:rq4}
Previous Meta-ICL methods rely on fixed-shot demonstrations during meta-training, limiting adaptability to diverse task complexities. MicLog overcomes this constraint through 0-shot to $k$-shot ProgMeta-ICL demonstrations.

As shown in Table \ref{tab:shotdiff}, MicLog utilizes meta-training data from 0-shot to 5-shot, while other variants rely solely on single-shot. All models are evaluated with 5-shot ICL prompts. Results indicate that LLMs meta-trained on any single shot (0/1/3/5) perform similarly across metrics and significantly outperform non-Meta-ICL baselines, showing that minimal meta-training substantially improves ICL utilization. Moreover, MicLog's progressive training yields further significant improvements, achieving over 90\% accuracy on two TA metrics. This demonstrates its superior ability to build in-context understanding.

\section{Conclusion}
In this paper, we introduce MicLog, an effective and efficient log parsing framework that boosts LLMs ICL capability via ProgMeta-ICL. MicLog employs a progressive meta-learning process (from 0-shot to few-shot), supported by a weighted DBSCAN algorithm for sampling highly representative examples.
Prior to LLM querying, a multi-level cache matches raw logs with existing templates and updates new templates from the LLM response on mismatches. Furthermore, we propose an efficient enhanced BM25 method for retrieving similar examples for prompt generation. Rigorous evaluation on benchmark datasets demonstrates that MicLog significantly outperforms existing parsers in both accuracy and robustness, highlighting its potential for log analysis research and practice.

\bibliography{aaai2026}

\begin{thebibliography}{51}
\providecommand{\natexlab}[1]{#1}

\bibitem[{Aky{\"u}rek et~al.(2022)Aky{\"u}rek, Schuurmans, Andreas, Ma, and Zhou}]{akyurek2022learning}
Aky{\"u}rek, E.; Schuurmans, D.; Andreas, J.; Ma, T.; and Zhou, D. 2022.
\newblock What learning algorithm is in-context learning? investigations with linear models.
\newblock \emph{arXiv preprint arXiv:2211.15661}.

\bibitem[{Amar and Rigby(2019)}]{amar2019mining}
Amar, A.; and Rigby, P.~C. 2019.
\newblock Mining historical test logs to predict bugs and localize faults in the test logs.
\newblock In \emph{2019 IEEE/ACM 41st International Conference on Software Engineering (ICSE)}, 140--151. IEEE.

\bibitem[{Bilgin et~al.(2020)Bilgin, Ersoy, Soykan, Tomur, {\c{C}}omak, and Kara{\c{c}}ay}]{bilgin2020vulnerability}
Bilgin, Z.; Ersoy, M.~A.; Soykan, E.~U.; Tomur, E.; {\c{C}}omak, P.; and Kara{\c{c}}ay, L. 2020.
\newblock Vulnerability prediction from source code using machine learning.
\newblock \emph{IEEE Access}, 8: 150672--150684.

\bibitem[{Brown(2020)}]{brown2020language}
Brown, T.~B. 2020.
\newblock Language models are few-shot learners.
\newblock \emph{arXiv preprint arXiv:2005.14165}.

\bibitem[{Chen et~al.(2021)Chen, Zhong, Zha, Karypis, and He}]{chen2021meta}
Chen, Y.; Zhong, R.; Zha, S.; Karypis, G.; and He, H. 2021.
\newblock Meta-learning via language model in-context tuning.
\newblock \emph{arXiv preprint arXiv:2110.07814}.

\bibitem[{Coda-Forno et~al.(2023)Coda-Forno, Binz, Akata, Botvinick, Wang, and Schulz}]{coda2023meta}
Coda-Forno, J.; Binz, M.; Akata, Z.; Botvinick, M.; Wang, J.; and Schulz, E. 2023.
\newblock Meta-in-context learning in large language models.
\newblock \emph{Advances in Neural Information Processing Systems}, 36: 65189--65201.

\bibitem[{Dai et~al.(2020)Dai, Li, Chen, Shang, and Chen}]{dai2020logram}
Dai, H.; Li, H.; Chen, C.-S.; Shang, W.; and Chen, T.-H. 2020.
\newblock Logram: Efficient log parsing using $ n $ n-gram dictionaries.
\newblock \emph{IEEE Transactions on Software Engineering}, 48(3): 879--892.

\bibitem[{Dong et~al.(2022)Dong, Li, Dai, Zheng, Ma, Li, Xia, Xu, Wu, Liu et~al.}]{dong2022survey}
Dong, Q.; Li, L.; Dai, D.; Zheng, C.; Ma, J.; Li, R.; Xia, H.; Xu, J.; Wu, Z.; Liu, T.; et~al. 2022.
\newblock A survey on in-context learning.
\newblock \emph{arXiv preprint arXiv:2301.00234}.

\bibitem[{Du and Li(2016)}]{du2016spell}
Du, M.; and Li, F. 2016.
\newblock Spell: Streaming parsing of system event logs.
\newblock In \emph{2016 IEEE 16th International Conference on Data Mining (ICDM)}, 859--864. IEEE.

\bibitem[{Du et~al.(2017)Du, Li, Zheng, and Srikumar}]{du2017deeplog}
Du, M.; Li, F.; Zheng, G.; and Srikumar, V. 2017.
\newblock Deeplog: Anomaly detection and diagnosis from system logs through deep learning.
\newblock In \emph{Proceedings of the 2017 ACM SIGSAC conference on computer and communications security}, 1285--1298.

\bibitem[{Ester et~al.(1996)Ester, Kriegel, Sander, Xu et~al.}]{ester1996density}
Ester, M.; Kriegel, H.-P.; Sander, J.; Xu, X.; et~al. 1996.
\newblock A density-based algorithm for discovering clusters in large spatial databases with noise.
\newblock In \emph{kdd}, volume~96, 226--231.

\bibitem[{Han et~al.(2017)Han, Li, Xing, Liu, and Feng}]{han2017learning}
Han, Z.; Li, X.; Xing, Z.; Liu, H.; and Feng, Z. 2017.
\newblock Learning to predict severity of software vulnerability using only vulnerability description.
\newblock In \emph{2017 IEEE International conference on software maintenance and evolution (ICSME)}, 125--136. IEEE.

\bibitem[{He et~al.(2017)He, Zhu, Zheng, and Lyu}]{he2017drain}
He, P.; Zhu, J.; Zheng, Z.; and Lyu, M.~R. 2017.
\newblock Drain: An online log parsing approach with fixed depth tree.
\newblock In \emph{2017 IEEE international conference on web services (ICWS)}, 33--40. IEEE.

\bibitem[{He et~al.(2021)He, He, Chen, Yang, Su, and Lyu}]{he2021survey}
He, S.; He, P.; Chen, Z.; Yang, T.; Su, Y.; and Lyu, M.~R. 2021.
\newblock A survey on automated log analysis for reliability engineering.
\newblock \emph{ACM computing surveys (CSUR)}, 54(6): 1--37.

\bibitem[{He et~al.(2018)He, Lin, Lou, Zhang, Lyu, and Zhang}]{he2018identifying}
He, S.; Lin, Q.; Lou, J.-G.; Zhang, H.; Lyu, M.~R.; and Zhang, D. 2018.
\newblock Identifying impactful service system problems via log analysis.
\newblock In \emph{Proceedings of the 2018 26th ACM joint meeting on European software engineering conference and symposium on the foundations of software engineering}, 60--70.

\bibitem[{He et~al.(2016)He, Zhu, He, and Lyu}]{he2016experience}
He, S.; Zhu, J.; He, P.; and Lyu, M.~R. 2016.
\newblock Experience report: System log analysis for anomaly detection.
\newblock In \emph{2016 IEEE 27th international symposium on software reliability engineering (ISSRE)}, 207--218. IEEE.

\bibitem[{Huang et~al.(2024{\natexlab{a}})Huang, Jiang, Chen, and Lyu}]{huang2024lunar}
Huang, J.; Jiang, Z.; Chen, Z.; and Lyu, M.~R. 2024{\natexlab{a}}.
\newblock LUNAR: Unsupervised LLM-based log parsing.
\newblock \emph{arXiv preprint arXiv:2406.07174}.

\bibitem[{Huang et~al.(2024{\natexlab{b}})Huang, Jiang, Liu, Huo, Gu, Chen, Feng, Dong, Yang, and Lyu}]{huang2024demystifying}
Huang, J.; Jiang, Z.; Liu, J.; Huo, Y.; Gu, J.; Chen, Z.; Feng, C.; Dong, H.; Yang, Z.; and Lyu, M.~R. 2024{\natexlab{b}}.
\newblock Demystifying and Extracting Fault-indicating Information from Logs for Failure Diagnosis.
\newblock In \emph{2024 IEEE 35th International Symposium on Software Reliability Engineering (ISSRE)}, 511--522. IEEE.

\bibitem[{Huo et~al.(2023)Huo, Su, Lee, and Lyu}]{huo2023semparser}
Huo, Y.; Su, Y.; Lee, C.; and Lyu, M.~R. 2023.
\newblock Semparser: A semantic parser for log analytics.
\newblock In \emph{2023 IEEE/ACM 45th International Conference on Software Engineering (ICSE)}, 881--893. IEEE.

\bibitem[{Jiang et~al.(2024{\natexlab{a}})Jiang, Liu, Chen, Li, Huang, Huo, He, Gu, and Lyu}]{jiang2024lilac}
Jiang, Z.; Liu, J.; Chen, Z.; Li, Y.; Huang, J.; Huo, Y.; He, P.; Gu, J.; and Lyu, M.~R. 2024{\natexlab{a}}.
\newblock LILAC: Log parsing using LLMs with adaptive parsing cache.
\newblock \emph{Proceedings of the ACM on Software Engineering}, 1(FSE): 137--160.

\bibitem[{Jiang et~al.(2024{\natexlab{b}})Jiang, Liu, Huang, Li, Huo, Gu, Chen, Zhu, and Lyu}]{jiang2024large}
Jiang, Z.; Liu, J.; Huang, J.; Li, Y.; Huo, Y.; Gu, J.; Chen, Z.; Zhu, J.; and Lyu, M.~R. 2024{\natexlab{b}}.
\newblock A large-scale evaluation for log parsing techniques: How far are we?
\newblock In \emph{Proceedings of the 33rd ACM SIGSOFT International Symposium on Software Testing and Analysis}, 223--234.

\bibitem[{Khan et~al.(2022)Khan, Shin, Bianculli, and Briand}]{khan2022guidelines}
Khan, Z.~A.; Shin, D.; Bianculli, D.; and Briand, L. 2022.
\newblock Guidelines for assessing the accuracy of log message template identification techniques.
\newblock In \emph{Proceedings of the 44th International Conference on Software Engineering}, 1095--1106.

\bibitem[{Le and Zhang(2023)}]{le2023log}
Le, V.-H.; and Zhang, H. 2023.
\newblock Log parsing with prompt-based few-shot learning.
\newblock In \emph{2023 IEEE/ACM 45th International Conference on Software Engineering (ICSE)}, 2438--2449. IEEE.

\bibitem[{Li, Wang, and Ke(2023)}]{li2023revisiting}
Li, G.; Wang, P.; and Ke, W. 2023.
\newblock Revisiting large language models as zero-shot relation extractors.
\newblock \emph{arXiv preprint arXiv:2310.05028}.

\bibitem[{Li et~al.(2024{\natexlab{a}})Li, Wang, Liu, Guo, Ji, Shang, and Xu}]{li2024meta}
Li, G.; Wang, P.; Liu, J.; Guo, Y.; Ji, K.; Shang, Z.; and Xu, Z. 2024{\natexlab{a}}.
\newblock Meta In-Context Learning Makes Large Language Models Better Zero and Few-Shot Relation Extractors.
\newblock \emph{arXiv preprint arXiv:2404.17807}.

\bibitem[{Li et~al.(2024{\natexlab{b}})Li, Fu, Huang, Yu, Li, Lai, and Ma}]{li2024revisiting}
Li, Z.; Fu, Q.; Huang, Z.; Yu, J.; Li, Y.; Lai, Y.; and Ma, Y. 2024{\natexlab{b}}.
\newblock Revisiting Log Parsing: The Present, the Future, and the Uncertainties.
\newblock \emph{IEEE Transactions on Reliability}.

\bibitem[{Lin et~al.(2016)Lin, Zhang, Lou, Zhang, and Chen}]{lin2016log}
Lin, Q.; Zhang, H.; Lou, J.-G.; Zhang, Y.; and Chen, X. 2016.
\newblock Log clustering based problem identification for online service systems.
\newblock In \emph{Proceedings of the 38th International Conference on Software Engineering Companion}, 102--111.

\bibitem[{Liu(2019)}]{liu2019roberta}
Liu, Y. 2019.
\newblock Roberta: A robustly optimized bert pretraining approach.
\newblock \emph{arXiv preprint arXiv:1907.11692}.

\bibitem[{Liu et~al.(2022)Liu, Zhang, He, Zhang, Li, Kang, Xu, Ma, Lin, Dang et~al.}]{liu2022uniparser}
Liu, Y.; Zhang, X.; He, S.; Zhang, H.; Li, L.; Kang, Y.; Xu, Y.; Ma, M.; Lin, Q.; Dang, Y.; et~al. 2022.
\newblock Uniparser: A unified log parser for heterogeneous log data.
\newblock In \emph{Proceedings of the ACM Web Conference 2022}, 1893--1901.

\bibitem[{Ma et~al.(2021)Ma, Zhou, Gui, Tan, Li, Zhang, and Huang}]{ma2021template}
Ma, R.; Zhou, X.; Gui, T.; Tan, Y.; Li, L.; Zhang, Q.; and Huang, X. 2021.
\newblock Template-free prompt tuning for few-shot NER.
\newblock \emph{arXiv preprint arXiv:2109.13532}.

\bibitem[{Ma, Kim, and Chen(2024)}]{ma2024librelog}
Ma, Z.; Kim, D.~J.; and Chen, T.-H. 2024.
\newblock LibreLog: Accurate and Efficient Unsupervised Log Parsing Using Open-Source Large Language Models.
\newblock \emph{arXiv preprint arXiv:2408.01585}.

\bibitem[{Min et~al.(2021)Min, Lewis, Zettlemoyer, and Hajishirzi}]{min2021metaicl}
Min, S.; Lewis, M.; Zettlemoyer, L.; and Hajishirzi, H. 2021.
\newblock Metaicl: Learning to learn in context.
\newblock \emph{arXiv preprint arXiv:2110.15943}.

\bibitem[{Min et~al.(2022)Min, Lyu, Holtzman, Artetxe, Lewis, Hajishirzi, and Zettlemoyer}]{min2022rethinking}
Min, S.; Lyu, X.; Holtzman, A.; Artetxe, M.; Lewis, M.; Hajishirzi, H.; and Zettlemoyer, L. 2022.
\newblock Rethinking the role of demonstrations: What makes in-context learning work?
\newblock \emph{arXiv preprint arXiv:2202.12837}.

\bibitem[{Rey and Neuh{\"a}user(2011)}]{rey2011wilcoxon}
Rey, D.; and Neuh{\"a}user, M. 2011.
\newblock Wilcoxon-signed-rank test.
\newblock In \emph{International encyclopedia of statistical science}, 1658--1659. Springer.

\bibitem[{Robertson(2004)}]{robertson2004understanding}
Robertson, S. 2004.
\newblock Understanding inverse document frequency: on theoretical arguments for IDF.
\newblock \emph{Journal of documentation}, 60(5): 503--520.

\bibitem[{Robertson, Zaragoza et~al.(2009)}]{robertson2009probabilistic}
Robertson, S.; Zaragoza, H.; et~al. 2009.
\newblock The probabilistic relevance framework: BM25 and beyond.
\newblock \emph{Foundations and Trends{\textregistered} in Information Retrieval}, 3(4): 333--389.

\bibitem[{Rubin, Herzig, and Berant(2021)}]{rubin2021learning}
Rubin, O.; Herzig, J.; and Berant, J. 2021.
\newblock Learning to retrieve prompts for in-context learning.
\newblock \emph{arXiv preprint arXiv:2112.08633}.

\bibitem[{Schubert et~al.(2017)Schubert, Sander, Ester, Kriegel, and Xu}]{schubert2017dbscan}
Schubert, E.; Sander, J.; Ester, M.; Kriegel, H.~P.; and Xu, X. 2017.
\newblock DBSCAN revisited, revisited: why and how you should (still) use DBSCAN.
\newblock \emph{ACM Transactions on Database Systems (TODS)}, 42(3): 1--21.

\bibitem[{Shima(2016)}]{shima2016length}
Shima, K. 2016.
\newblock Length matters: Clustering system log messages using length of words.
\newblock \emph{arXiv preprint arXiv:1611.03213}.

\bibitem[{Wu, Yu, and Li(2024)}]{wu2024log}
Wu, Y.; Yu, S.; and Li, Y. 2024.
\newblock Log Parsing using LLMs with Self-Generated In-Context Learning and Self-Correction.
\newblock \emph{arXiv preprint arXiv:2406.03376}.

\bibitem[{Xiao, Le, and Zhang(2024)}]{xiao2024free}
Xiao, Y.; Le, V.-H.; and Zhang, H. 2024.
\newblock Free: Towards More Practical Log Parsing with Large Language Models.
\newblock In \emph{Proceedings of the 39th IEEE/ACM International Conference on Automated Software Engineering}, 153--165.

\bibitem[{Xu et~al.(2024{\natexlab{a}})Xu, Cui, Zhao, Zhang, He, He, Li, Kang, Lin, Dang et~al.}]{xu2024unilog}
Xu, J.; Cui, Z.; Zhao, Y.; Zhang, X.; He, S.; He, P.; Li, L.; Kang, Y.; Lin, Q.; Dang, Y.; et~al. 2024{\natexlab{a}}.
\newblock Unilog: Automatic logging via llm and in-context learning.
\newblock In \emph{Proceedings of the 46th ieee/acm international conference on software engineering}, 1--12.

\bibitem[{Xu et~al.(2024{\natexlab{b}})Xu, Yang, Huo, Zhang, and He}]{xu2024divlog}
Xu, J.; Yang, R.; Huo, Y.; Zhang, C.; and He, P. 2024{\natexlab{b}}.
\newblock DivLog: Log Parsing with Prompt Enhanced In-Context Learning.
\newblock In \emph{Proceedings of the IEEE/ACM 46th International Conference on Software Engineering}, 1--12.

\bibitem[{Yang et~al.(2024)Yang, Yang, Hui, Zheng, Yu, Zhou, Li, Li, Liu, Huang, Dong, Wei, Lin, Tang, Wang, Yang, Tu, Zhang, Ma, Xu, Zhou, Bai, He, Lin, Dang, Lu, Chen, Yang, Li, Xue, Ni, Zhang, Wang, Peng, Men, Gao, Lin, Wang, Bai, Tan, Zhu, Li, Liu, Ge, Deng, Zhou, Ren, Zhang, Wei, Ren, Fan, Yao, Zhang, Wan, Chu, Liu, Cui, Zhang, and Fan}]{qwen2}
Yang, A.; Yang, B.; Hui, B.; Zheng, B.; Yu, B.; Zhou, C.; Li, C.; Li, C.; Liu, D.; Huang, F.; Dong, G.; Wei, H.; Lin, H.; Tang, J.; Wang, J.; Yang, J.; Tu, J.; Zhang, J.; Ma, J.; Xu, J.; Zhou, J.; Bai, J.; He, J.; Lin, J.; Dang, K.; Lu, K.; Chen, K.; Yang, K.; Li, M.; Xue, M.; Ni, N.; Zhang, P.; Wang, P.; Peng, R.; Men, R.; Gao, R.; Lin, R.; Wang, S.; Bai, S.; Tan, S.; Zhu, T.; Li, T.; Liu, T.; Ge, W.; Deng, X.; Zhou, X.; Ren, X.; Zhang, X.; Wei, X.; Ren, X.; Fan, Y.; Yao, Y.; Zhang, Y.; Wan, Y.; Chu, Y.; Liu, Y.; Cui, Z.; Zhang, Z.; and Fan, Z. 2024.
\newblock Qwen2 Technical Report.
\newblock \emph{arXiv preprint arXiv:2407.10671}.

\bibitem[{Yao et~al.(2024)Yao, Duan, Xu, Cai, Sun, and Zhang}]{yao2024survey}
Yao, Y.; Duan, J.; Xu, K.; Cai, Y.; Sun, Z.; and Zhang, Y. 2024.
\newblock A survey on large language model (llm) security and privacy: The good, the bad, and the ugly.
\newblock \emph{High-Confidence Computing}, 100211.

\bibitem[{Yu et~al.(2023)Yu, He, Chen, and Wu}]{yu2023brain}
Yu, S.; He, P.; Chen, N.; and Wu, Y. 2023.
\newblock Brain: Log parsing with bidirectional parallel tree.
\newblock \emph{IEEE Transactions on Services Computing}, 16(5): 3224--3237.

\bibitem[{Zhang et~al.(2019)Zhang, Xu, Lin, Qiao, Zhang, Dang, Xie, Yang, Cheng, Li et~al.}]{zhang2019robust}
Zhang, X.; Xu, Y.; Lin, Q.; Qiao, B.; Zhang, H.; Dang, Y.; Xie, C.; Yang, X.; Cheng, Q.; Li, Z.; et~al. 2019.
\newblock Robust log-based anomaly detection on unstable log data.
\newblock In \emph{Proceedings of the 2019 27th ACM joint meeting on European software engineering conference and symposium on the foundations of software engineering}, 807--817.

\bibitem[{Zhang et~al.(2023)Zhang, Zhang, Yang, and Wang}]{zhang2023and}
Zhang, Y.; Zhang, F.; Yang, Z.; and Wang, Z. 2023.
\newblock What and how does in-context learning learn? bayesian model averaging, parameterization, and generalization.
\newblock \emph{arXiv preprint arXiv:2305.19420}.

\bibitem[{Zhao et~al.(2023)Zhao, Zhou, Li, Tang, Wang, Hou, Min, Zhang, Zhang, Dong et~al.}]{zhao2023survey}
Zhao, W.~X.; Zhou, K.; Li, J.; Tang, T.; Wang, X.; Hou, Y.; Min, Y.; Zhang, B.; Zhang, J.; Dong, Z.; et~al. 2023.
\newblock A survey of large language models.
\newblock \emph{arXiv preprint arXiv:2303.18223}, 1(2).

\bibitem[{Zhong et~al.(2024)Zhong, Mo, Liu, Liu, Lu, Zhou, Wu, Li, and Wen}]{zhong2024logparser}
Zhong, A.; Mo, D.; Liu, G.; Liu, J.; Lu, Q.; Zhou, Q.; Wu, J.; Li, Q.; and Wen, Q. 2024.
\newblock LogParser-LLM: Advancing Efficient Log Parsing with Large Language Models.
\newblock In \emph{Proceedings of the 30th ACM SIGKDD Conference on Knowledge Discovery and Data Mining}, 4559--4570.

\bibitem[{Zhu et~al.(2019)Zhu, He, Liu, He, Xie, Zheng, and Lyu}]{zhu2019tools}
Zhu, J.; He, S.; Liu, J.; He, P.; Xie, Q.; Zheng, Z.; and Lyu, M.~R. 2019.
\newblock Tools and benchmarks for automated log parsing.
\newblock In \emph{2019 IEEE/ACM 41st International Conference on Software Engineering: Software Engineering in Practice (ICSE-SEIP)}, 121--130. IEEE.

\end{thebibliography}

\end{document}